\documentclass[a4paper]{IEEEtran}
\IEEEoverridecommandlockouts

\usepackage{cite}
\usepackage{amsmath,amssymb,amsfonts}
\usepackage{algorithmic}
\usepackage{graphicx}
\usepackage{textcomp}
\usepackage{multirow}
\usepackage{svg}
\usepackage{fancyhdr}
\usepackage{epstopdf}
\usepackage{paralist}
\usepackage{url}
\usepackage{lipsum}
\usepackage{xcolor,colortbl}
\usepackage{mathtools}
\usepackage{color}
\usepackage{eurosym} 
\usepackage{booktabs}
\usepackage{tikz}
\usepackage[super]{nth}
\usepackage{comment}

\usepackage{verbatim} 
\usepackage{enumerate}
\usepackage{subfigure}
\usepackage{caption}
\usepackage{xurl}
\usepackage{hyperref}
\usepackage{comment}
\usepackage[commandnameprefix=always]{changes}

\usepackage[shortlabels]{enumitem}

\def\BibTeX{{\rm B\kern-.05em{\sc i\kern-.025em b}\kern-.08em
    T\kern-.1667em\lower.7ex\hbox{E}\kern-.125emX}}
\begin{document}

\title{AI-Native Orchestration in the 6G Continuum: Evolving Operator Platforms with Agentic AI
	\thanks{C. Carballo González, H. Chergui, S. Giménez-Antón, M. Mosahebfard, J. S. Camargo, and P. Sayyad Khodashenas are with the i2CAT Foundation, 08034 Barcelona, Spain (e-mails: {claudia.carballo, hatim.chergui, sergio.gimenez, mohammadreza.mosahebfard, juan.camargo, pouria.khodashenas}@i2cat.net). V. Theodorou is with Intracom Telecom (e-mail: theovas@intracom-telecom.com). C. Verikoukis is with ISI-ATH and University of Patras, Greece (e-mail: chverik@isi.gr).}
	\thanks{This work has been co-funded by the European Union under the SUNRISE-6G project (Grant Agreement No. 101139257). It has also been supported by the COALESCE-6G project (PID2024-163028OB-I00), funded by MICIU/AEI/10.13039/501100011033/FEDER, EU. (Grant Agreement No. 101139257).}}

\author{
\IEEEauthorblockN{
Claudia Carballo Gonz\'{a}lez\textsuperscript{1},
Hatim Chergui\textsuperscript{1},
Sergio Gim\'{e}nez-Ant\'{o}n\textsuperscript{1},
Mohammadreza Mosahebfard\textsuperscript{1},\\
Juan Sebasti\'{a}n Camargo\textsuperscript{1},
Pouria Sayyad Khodashenas\textsuperscript{1},
Vasileios Theodorou\textsuperscript{2},
and Christos Verikoukis\textsuperscript{3}\\
}
\vspace{0.2cm}
\IEEEauthorblockA{
	\textsuperscript{1}i2CAT Foundation, Barcelona, Spain;
}
\IEEEauthorblockA{
	\textsuperscript{2}Intracom Telecom, 
    Greece;\\
}
\vspace{0.1cm}
\IEEEauthorblockA{
	\textsuperscript{3}Industrial Systems Institute (ISI)/Athena Research Center and University of Patras, 
    Greece\\
}
}
\maketitle

\begin{abstract}
 
As Sixth-Generation (6G) networks evolve towards a seamless Cloud-Edge-Internet of Things (IoT) continuum, autonomous orchestration across distributed compute and network environments becomes critical. Future 6G services are expected to span multiple administrative and operator domains, making federation essential for enabling ubiquitous, ultra-low-latency service continuity beyond individual footprint limits. This operational complexity demands AI-native mechanisms that support intent-driven automation and closed-loop management. While the GSMA Operator Platform (OP) provides the architectural blueprint for multi-operator federation and network capability exposure, and the ETSI Software Development Group OpenOP (SDG OOP) offers a primary open-source reference implementation of this blueprint, current frameworks are limited by stateless API interactions and lack native intelligence. This paper proposes an Agentic-driven Intelligence extension for the GSMA OP architecture, using the OOP as the reference framework. We introduce an AI-native orchestration layer that leverages autonomous agents to manage persistent service contexts and enable closed-loop control via CAMARA APIs. By integrating a Declarative Monitoring and Alerting System (DeMAS) into the OOP stack and establishing a decentralised agent negotiation protocol, the proposed architecture enables real-time, intent-driven resource optimisation and autonomous cross-domain conflict resolution across federated domains. We validate our approach through a representative 6G use case involving Ultra-Reliable Low-Latency Communication (URLLC) and enhanced Mobile Broadband (eMBB) coexistence, demonstrating that an agentic OP framework autonomously reconciles stringent Service Level Agreements (SLAs) while enhancing infrastructure energy efficiency. Our findings establish a scalable blueprint for future cross-domain Network-as-a-Service (NaaS) models that align standardised exposure with 6G autonomous requirements.
\end{abstract}

\section{Introduction}
\label{sec:intro}



The transition towards Sixth-Generation (6G) networks, together with the emergence of the Cloud-Edge-Internet of Things (IoT) 
continuum, is significantly increasing the complexity of managing ubiquitous connectivity and stringent service requirements. 
Future 6G services are expected to seamlessly extend this continuum across multiple administrative and network domains, requiring interoperable mechanisms for end-to-end (E2E) service continuity and coordinated resource management. However, existing mobile networks and Multi-access Edge Computing (MEC) deployments remain highly fragmented, typically operating within isolated operator domains~\cite{dalgitsis20266g}, which significantly complicates developer integration and limits support for next-generation latency-sensitive applications. 
To address these challenges, the adoption of the “3C network” vision~\cite{11037024} is becoming essential, enabling the seamless convergence and coordinated orchestration of communication (network resources), computation (edge-cloud resources), and control (autonomous closed-loop mechanisms) across federated, multi-domain ecosystems. 

The GSMA Open Gateway (OGW) initiative and the Operator Platform (OP) architecture are key enablers for overcoming  these barriers  and establishing an interoperable landscape~\cite{pino2025demonstrating, wang2025global}.
Notably, 86 operator groups, representing more than 300 networks and 80\% of global mobile connections are aligned around this ecosystem~\cite{gsma_news}. This vision is being realised through open-source initiatives such as the Linux Foundation’s CAMARA Project~\cite{CAMARA_API_Overview}, transforming fragmented telecommunications infrastructure into a cohesive programmable platform by exposing capabilities through operator-neutral Application Programming Interfaces (APIs). In combination with the TM Forum (TMF)’s Open Digital Architecture for standardised Business and Operational Support Systems (BSS/OSS), they facilitate multi-operator federation and the seamless realisation of cross-domain Network-as-a-Service (NaaS) models. Complementing this ecosystem, the ETSI Software Development Group OpenOP (SDG OOP) is actively developing an open-source reference implementation of the GSMA OP to accelerate innovation and market adoption by bridging standardisation, research and industrial communities~\cite{sdg_oop}. 

Despite these advancements, the transition towards fully autonomous 6G networks requires integrating Artificial Intelligence (AI) to evolve from an external add-on into a native system capability for telecom-specific applications~\cite{brik2024explainable, chen2025toward}. In this context, agentic AI (i.e., autonomous agents capable of planning tasks, reasoning across data sources, and interacting with external systems) is emerging as a key enabler for reshaping how network APIs are exposed, discovered, and consumed~\cite{gsma_news}. Achieving this level of autonomy, aligned with targets such as TMF’s Levels 4 and 5 (closed-loop and fully autonomous operation), requires moving beyond conventional automation paradigms towards agent-driven systems~\cite{coronado2025ai}.

However, AI integration within the OP framework is still at an early stage. An initial attempt in this direction is the recent OOP Release 1, published in early 2026, which incorporates an AI Integration ($\text{AI}^2$) module that exposes a Model Context Protocol (MCP) server and a lightweight AI agent for intent-based interaction with OP northbound capabilities. While this deployment marks a step forward, we position the OOP Release~1 $\text{AI}^2$ module as the current baseline: a stateless, single-agent north-to-south translator that maps isolated natural-language prompts to CAMARA API operations within a single OP instance. These early-stage approaches lack the persistent context and multi-agent coordination required for complex reasoning, especially when managing conflicting service requirements across multiple operator domains. Furthermore, current OP implementations, including OOP, do not yet provide real-time observability, integration of Network Digital Twins (NDTs), or decentralised negotiation protocols necessary to support truly autonomous, closed-loop, and intent-driven network control.

To address these challenges, this paper proposes an Agentic-driven Intelligence module that evolves the current OP blueprint into an AI-native ecosystem for a truly connected and collaborative federated 6G framework. Our novelty, therefore, starts upstream of the Release~1 $\text{AI}^2$ baseline: instead of another prompt-to-API interface, we introduce an Agentic-driven Intelligence layer for stateful, peer-to-peer, decentralised coordination across federated operator domains. While preliminary implementations in frameworks such as ETSI SDG OOP have explored initial AI integration, the proposed architecture explicitly extends that single-domain baseline by introducing an autonomous Agentic Negotiation Layer that leverages 
the Agent-to-Agent (A2A) protocol for cross-domain resource conflict resolution. By grounding agent decisions in NDTs and a real-time observability framework, we enable a highly responsive closed-loop system. We also integrate TMF APIs for intent management, metrics and alarm propagation, alongside CAMARA APIs to natively manage converged compute and network resources, effectively transforming static network APIs into intelligent and self-optimising building blocks.

The main contributions of this article are summarised as follows:
\begin{itemize}

\item An architectural extension of the GSMA OP blueprint, using the ETSI SDG OOP as the reference framework, through an Agentic-driven Intelligence module that integrates generative AI entities and context-aware interaction mechanisms across federated domains. This module leverages a peer-to-peer A2A negotiation protocol to enable autonomous resource coordination and conflict resolution among competing services, while utilising NDTs to ground agent reasoning through predictive simulations. 
\item The implementation framework of a Declarative Monitoring and Alerting System (DeMAS) aligned with CAMARA APIs and integrated with the Agentic-driven Intelligence module. This component provides real-time telemetry and persistent event context via an MCP configuration, enabling closed-loop, intent-driven network automation. 
\end{itemize}

The remainder of this paper is structured as follows. Section~\ref{sec:overview} provides a comprehensive overview of the GSMA OP architecture and the ETSI SDG OOP.  
Building on this, Section~\ref{sec:enhancedsol} presents the proposed enhanced OP framework and its agentic solution, using the ETSI SDG OOP as the reference implementation. A representative 6G use case demonstrating how the enhanced OP framework enables autonomous management of edge-cloud resources and network slice reconfiguration 
is presented 
in Section~\ref{sec:use-case}. This is followed by a discussion of related open research challenges in Section~\ref{sec:next}, before concluding the paper in Section~\ref{sec:conclusion}.

\section{The GSMA OGW/OP and SDG OOP Overview}
\label{sec:overview}

\begin{figure}[t]
	\centering
	\includegraphics[width=1.0\linewidth]{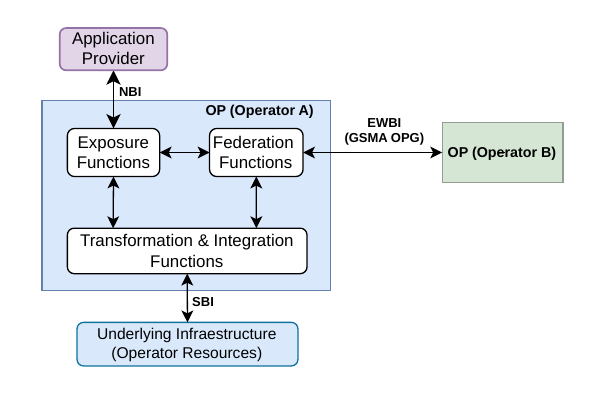}
	\caption{High-level GSMA OP architecture}
	\label{fig:op_high_level}
\end{figure}

\begin{figure*}[t]
	\centering
	  \includegraphics[width=0.92\linewidth]{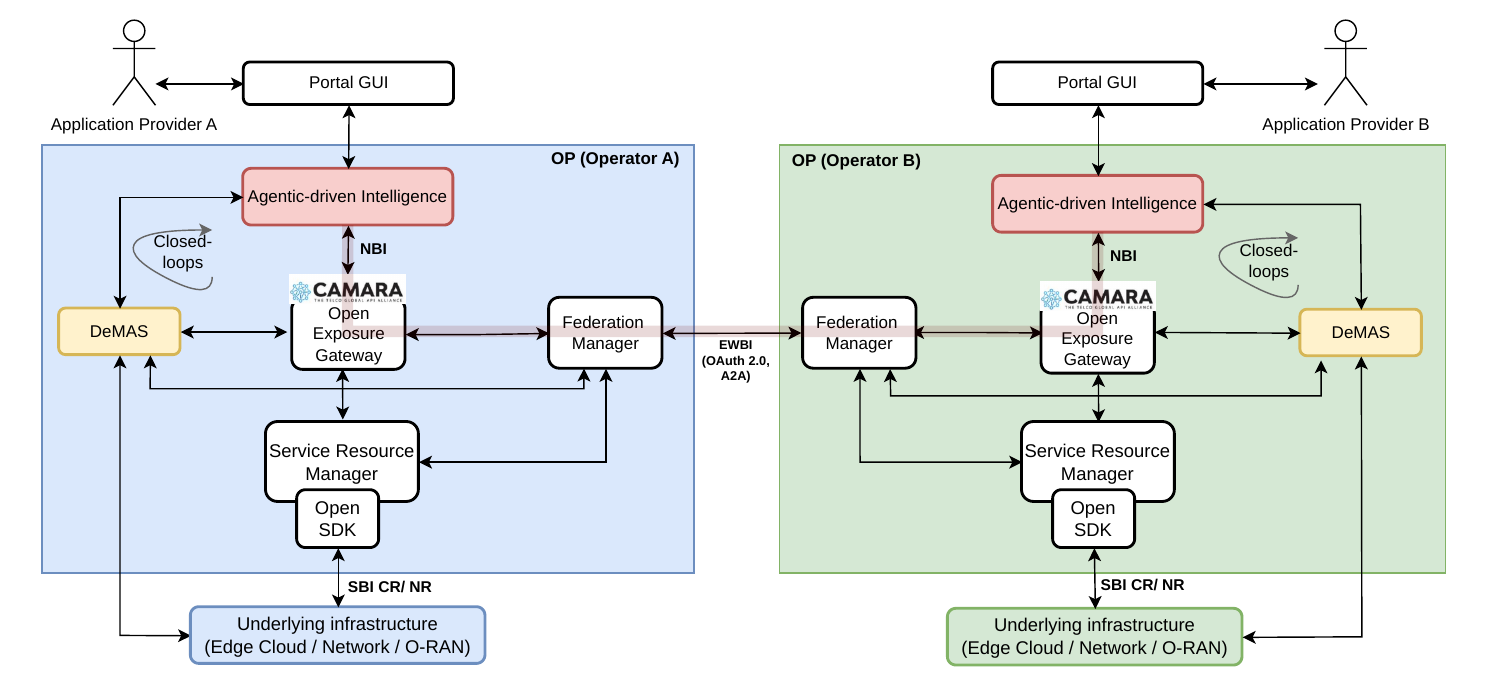}
	\caption{The enhanced OOP architecture}
	\label{fig:ext_oop}
\end{figure*}

The GSMA drives industry-wide collaboration and policy guidance to enable a connected, innovation-friendly mobile ecosystem, laying the technical foundations required for federated 6G networks~\cite{wang2025global}. Within this landscape, the GSMA OGW initiative plays a pivotal role by establishing a standardised framework of common network APIs, designed to provide developers with universal, operator-neutral access to telecommunications capabilities~\cite{gsma_ogw}. Specifically, it promotes the adoption of Service APIs (e.g., CAMARA) to abstract underlying network complexity for application developers, while simultaneously leveraging TMF Operate APIs to provide the business logic and operational layer for cross-carrier federation.

The GSMA OP~\cite{gsma_opg02_2026} serves as the technical foundation for the GSMA OGW, providing the standardised architecture required to expose network capabilities through a connect-once, connect-to-many model. By unifying service exposure and defining a robust federation interface, the OP allows operators to expand their reach across multi-domain ecosystems, enabling the seamless deployment of OGW APIs in collaboration with hyperscalers, telco, and edge-cloud providers.
As depicted in Fig.~\ref{fig:op_high_level}, the GSMA OP architecture is organised into distinct functional components~\cite{gsma_opg02_2026} that together manage the E2E service lifecycle across federated infrastructures:

\paragraph{Exposure Functions}
This level acts as the primary gateway for Application Providers and Aggregators. It manages the Northbound Interface (NBI), handling request termination and ensuring representational consistency so developers can work with uniform APIs regardless of the backing resources.

\paragraph{Federation Functions}
These functions manage the East-Westbound Interface (EWBI) to securely interconnect independent OP instances, facilitating the exchange of availability zone information and resource catalogues to extend service reach into federated partner footprints.

\paragraph{Transformation \& Integration Functions}
These functions serve as the logical processing engine and gateway to the Southbound Interface (SBI), mapping high-level NBI or EWBI requests into specific technical parameters while managing physical and protocol-specific connections via specialised interfaces for network (SBI-NR) and edge-cloud (SBI-CR) resource management.

The ETSI SDG OOP is an open-source implementation of the GSMA OP~\cite{sdg_oop}, fostering an interoperable 6G ecosystem for developers, researchers, and industry players. In this framework, the Open Exposure Gateway (OEG) implements the Exposure capabilities, allowing the unified exposure of testbed resources via CAMARA APIs. The Federation Manager (FM) is the interoperability brain of this framework, orchestrating the entire partnership between OPs. It enables a trusted connection, allowing developers to seamlessly deploy and manage their applications across multiple federated networks. Finally, the Transformation and Integration functions have been realised by developing the Service Resource Manager (SRM) and the Open Software Development Kit (SDK). The SRM manages resource allocation and interacts with the underlying infrastructure consuming the Open SDK, which packages several transformation functions to map standardised APIs, such as CAMARA, to platform-specific APIs for managing heterogeneous resources (e.g., Open Radio Access Network (O-RAN), edge-cloud platforms, Kubernetes clusters).
This framework also introduces a Portal, a web-based interface that provides operators, developers, and application providers with an intuitive graphical user interface (GUI) to interact with the OOP.

Starting from Release~1, OOP includes the $\text{AI}^2$ module, which combines an MCP server and a lightweight AI agent to expose OEG northbound capabilities through natural language and tool-based invocation. Architecturally, this module acts as a stateless assistant 
that translates prompts into chained CAMARA API calls but remains confined to a single operator without maintaining cross-domain state. Consequently, this baseline lacks the persistent context and live observability required for federated 6G environments. 

Crucially, a single-agent architecture introduces prohibitive operational complexity, as a solitary entity cannot efficiently process heterogeneous telemetry or manage multi-variable decision-making simultaneously. Instead, an intra-domain scenario requires a multi-agent framework where specialised entities manage distinct services or operational tasks (e.g., network and compute provisioning), balancing conflicting objective functions over finite infrastructure resources. Furthermore, these domain-specific agents must cooperatively negotiate both locally and across boundaries (inter-domain) to dynamically reconcile multi-dimensional resource parameters, guaranteeing seamless Service Level Agreement (SLA) continuity as mobile users traverse federated footprints. 

Building on these foundations, the following Section details our proposed Agentic-driven Intelligence and observability extensions, engineered to handle the rigorous autonomous and multi-domain requirements of the 6G era. Rather than replacing the OOP Release~1 $\text{AI}^2$ capability, the proposed Agentic-driven Intelligence layer is conceived as an upstream evolution focused on decentralised, multi-operator resource conflict resolution. By adopting the ETSI SDG OOP as the reference framework, ensuring full alignment with GSMA OP and CAMARA specifications, this approach evolves existing capabilities while introducing new, essential components into a fully closed-loop, multi-agent system. 

\section{Extending the OP Architecture Blueprint}
\label{sec:enhancedsol}

Looking ahead, the next step for programmable networks is not just the exposure of APIs, but the intelligence layered on top of them. Agentic AI can transform static APIs into self-optimising, goal-driven building blocks, enabling automatic discovery, orchestration, and secure access across operators. This reduces development complexity, accelerates service deployment, and unlocks new use cases ranging from adaptive media pipelines to autonomous IoT systems. GSMA OP provides the foundation, while agentic AI acts as the catalyst towards scalable, fully autonomous 6G networks.

Fig.~\ref{fig:ext_oop} illustrates our vision of an extended OP architecture, based on the open-source implementation of the ETSI SDG OOP, evolving towards a truly connected and collaborative federated 6G framework. This evolution combines Agentic-driven Intelligence with DeMAS, enabling control closed-loops across the telco–managed Cloud-Edge-IoT continuum. 

DeMAS integrates directly with the underlying heterogeneous infrastructure, enabling the collection of multi-dimensional observability data from diverse, platform-specific
sources. Furthermore, it interfaces with the FM to support the secure exchange across federated operators via the EWBI, providing a consistent, cross-layer view of system state that forms the basis for agent reasoning. To drive this automation, DeMAS integrates with Connectivity Insights and Application Profile APIs (DeMAS-OEG integration), exposing network and compute performance metrics and application-specific requirements. These requirements are dynamically registered
as predefined rules within the monitoring subsystem. If any operational parameter or requirement is unmet, DeMAS triggers targeted alerts. Both these active alerts and the underlying time-series metrics are streamed to the agentic layer, equipping the autonomous agents with the semantic and granular context needed for multi-agent coordination and decision-making.

Multiple agents are deployed at each OP domain to solve specialised tasks, leveraging the A2A protocol to resolve resource conflicts locally (intra-domain) and extend their cooperative negotiation globally (inter-domain) by encapsulating A2A messages across the EWBI. Through this multi-level negotiation framework, federated systems can dynamically orchestrate resources and applications E2E. For instance, specialised agents across different operators negotiate and subsequently invoke several APIs (e.g., CAMARA Quality on Demand (QoD) or Optimal Edge Discovery) to ensure an application is deployed at the edge node closest to the users while continuously optimising shared network and computational resources. 
To interface with this intelligence, the Portal GUI lets Application Providers submit intent-driven requests (e.g., deploying applications or requesting time-bounded QoS sessions) and visualise real-time monitoring status, ensuring both infrastructure efficiency and optimal user experience.

\subsection{Monitoring and alerting framework}
\label{sec:DeMAS_description}

DeMAS simplifies observability across distributed operator infrastructures. It abstracts the complexity of the underlying infrastructure, providing a single API endpoint through which applications can create and activate alerts dynamically on a Prometheus-based monitoring subsystem. Unlike general-purpose observability platforms, DeMAS is conceived as an API-first and standards-aligned abstraction layer for federated OPs, combining dynamic rule provisioning with subscription-based, event-driven alarm dissemination across domains. Using the Portal GUI, developers and infrastructure owners can define monitoring rules and thresholds for metrics ranging from compute, storage, and network usage to application-level QoS and quality of experience indicators. By centralising these capabilities, the extended OP framework reduces manual configuration, ensures a consistent monitoring experience, and streamlines observability across heterogeneous and federated environments.

\begin{figure}[t]
	\centering
	\includegraphics[width=0.93\linewidth]{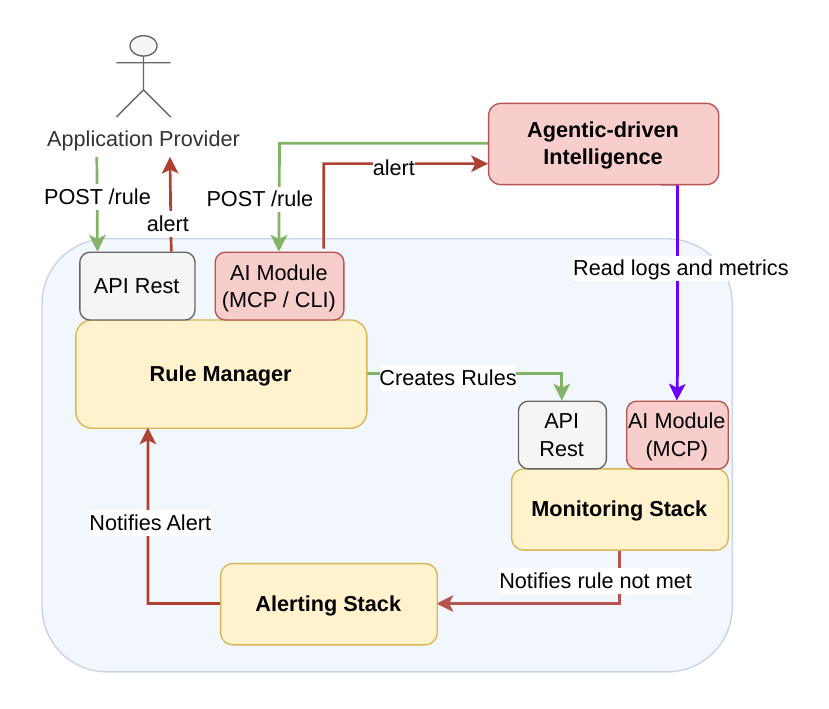}
	\caption{DeMAS framework}
	\label{fig:demas}
\end{figure}

Fig.~\ref{fig:demas} illustrates the internal architecture of DeMAS, highlighting its three core components: the Rule Manager, which exposes the API and handles rule evaluation; the Monitoring Stack (Prometheus/Thanos subsystem), enabling scalable metric collection, aggregation, deduplication, and long-term storage across distributed infrastructures; and the Alerting Stack, which processes triggered alerts and forwards them to Partner OPs. The TMF 642 API is integrated on top of this workflow, enabling standardised alarm propagation across federated operators via the EWBI while preserving operational consistency and compliance with GSMA~\cite{gsma_opg04_2025}. This modular architecture provides E2E visibility of resource and application performance, allowing Originating and Partner OPs to align on events, alarms, and metrics in real time while hiding underlying infrastructure complexity. By unifying monitoring 
and alarm propagation in a declarative, API-driven manner, DeMAS effectively strengthens operational resilience and guarantees consistent, cross-layer observability across federated 6G deployments.

\subsection{Agentic-driven intelligence framework}
\label{sec:AI_description}

The proposed autonomous resource management framework implements a multi-layered, closed-loop orchestration architecture that seamlessly traverses from high-level business intents to physical resource enforcement. This holistic approach is essential for supporting diverse and demanding 6G ecosystems, efficiently handling the competing requirements of advanced use cases such as Ultra-Reliable Low-Latency Communication (URLLC) and enhanced Mobile Broadband (eMBB)~\cite{singh2025towards}. 

Compared with the lightweight policy networks evaluated in recent works~\cite{bandara2026agentic}, language-based agents introduce additional deployment requirements in an edge-cloud-native environment. In the proposed framework, each agent is deployed as a Pod, the smallest deployable unit in a containerised Kubernetes cluster, and is attached to a sidecar MCP server that exposes DeMAS telemetry and NDT views, ensuring that the agent's context window is populated with task-relevant states rather than raw observations. 

\begin{figure}
	\centering
    \includegraphics[width=1.0\linewidth]{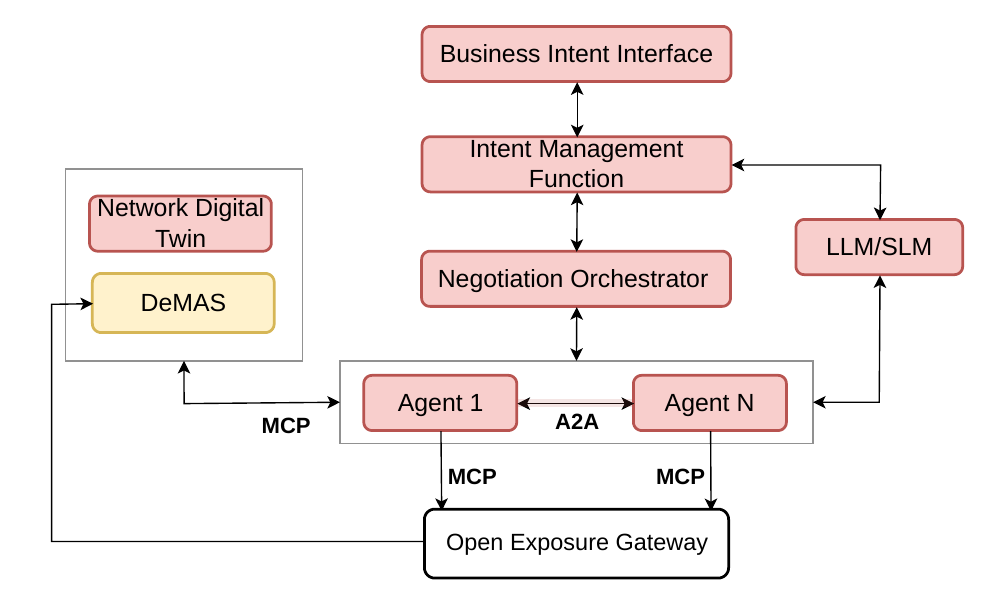}
	\caption{Agentic-driven Intelligence framework}
	\label{fig:agentic}
\end{figure}

Fig.~\ref{fig:agentic} depicts the proposed Agentic-driven Intelligence framework with its main components. Structurally, the operational lifecycle and interactions of these components are analysed across the following core layers:

\subsubsection{Intent Management and Agentic Negotiation Layer}
The orchestration lifecycle is initiated by the Intent Management Function (IMF), which aligns with standardised telecommunications ontologies, such as the TMF guidelines (e.g., TMF 921), to parse natural language business objectives into technical SLAs and quantifiable Service Level Objectives. Once translation and feasibility guardrails are applied to establish the theoretical performance floors of the underlying physical network, the framework transitions into the Agentic Negotiation Layer.

In this layer, autonomous, domain-specific agents across federated sites, engage in decentralised negotiation to allocate shared multidimensional resources, such as RAN throughput and compute capacity. To meet the stringent real-time requirements of 6G, the framework adopts a hybrid strategy: Small Language Model (SLM) instances are scheduled close to the data sources at the edge for rapid negotiation rounds, while heavier LLM workloads are hosted in centralised endpoints for complex, multi-domain strategy refinement~\cite{chen2025toward}.

To facilitate seamless interoperability and context-sharing, the agents utilise MCP 
for standardised, real-time access to operational telemetry. NDTs serve as the critical validation engine, ensuring reasoning is strictly grounded in the network's current state. By leveraging NDTs via the MCP-enabled sidecar, agents execute probabilistic forecasting (e.g., Monte Carlo simulations) for granular risk assessment.

The robust A2A communication protocol drives the negotiation process. Depending on the criticality of the supported service, agents dynamically adopt distinct reasoning paradigms: a Risk-Neglect approach focused on optimising average-case utility, or a Risk-Aware methodology, such as targeting the Conditional Value at Risk (CVaR), to guarantee extreme reliability bounds. The A2A negotiation unfolds through a multi-round alternating offer protocol, where agents in a single or multi-domain federated sites, iteratively exchange and evaluate \textit{Safe}, \textit{Aggressive}, and \textit{Balanced} proposals. 
These proposals are continuously scored using a utility function that balances SLA compliance with overarching operational constraints, including energy cost models and resource availability.

\subsubsection{API Enforcement Layer}
Upon reaching a consensus in the A2A negotiation phase, the framework propagates the agreed resource distribution down to the API Enforcement Layer. This layer utilises a low-latency message bus (e.g., Apache Kafka) to distribute agent decisions asynchronously, effectively decoupling high-level reasoning workflows from the operational constraints of the underlying infrastructure. 
For example, it interfaces with programmable Service APIs (e.g., CAMARA QoD) to enforce radio bandwidth prioritisation, invoke MEC and GSMA-aligned Service APIs for compute reservation across federated OP domains, and execute Kubernetes-native operations such as \texttt{ResourceQuota} updates or Pod rescheduling to optimise resource utilisation across the edge-cloud continuum.

Operating at a fine-grained temporal resolution, the environment continuously loops network feedback into the reasoning engine. When the NDT confirms high local stability, as indicated by low statistical variance in the network telemetry, the framework allows for complex, dynamic resource concessions among agents in intra- and inter-domain scenarios. Ultimately, the resource allocation logic abstracts underlying compute, 
network, and operational parameters into a generalised optimisation problem. This allows autonomous agents to balance targeted performance metrics, such as trade-offs between operational energy efficiency and stringent reliability guarantees, strictly in accordance with the enforced business intents during the final physical enforcement phase.

\section{Use Case Scenario: V2X and eMBB Coexistence}
\label{sec:use-case}

This Section instantiates the proposed Agentic-driven Intelligence framework by coupling DeMAS observability with autonomous agentic control to manage a critical 6G coexistence scenario. The orchestration lifecycle begins when Application Providers submit high-level service requirements using natural language through the Portal GUI. These intents are subsequently translated into declarative monitoring rules that capture, in real time, latency and network resource conditions. The resulting telemetry and alarms are directly ingested by the autonomous agents, providing the necessary live context to trigger dynamic, corrective actions that mitigate SLA degradation. By applying this architecture to the coexistence of safety-critical Vehicle-to-Everything (V2X) traffic and eMBB services, we demonstrate how the enhanced OP framework enables agents to take dynamic actions across a federated OP ecosystem. Specifically, these coexisting services are modeled as independent, virtual network slices competing for infrastructure resources. This coordinated approach ensures an optimised 
QoS regardless of the subscriber's location, while simultaneously maximising infrastructure energy efficiency.

\subsection{Scenario Setup and Agentic Logic}

In this scenario, the IMF translates business requirements into specific SLAs, such as a strict $\approx$ 10~ms latency requirement for V2X (URLLC), while enforcing feasibility guardrails against physical network floors ($L_{min} \approx$ 3~ms). During the Agentic Negotiation Layer phase, autonomous service-specific agents (i.e., a URLLC agent dedicated to safety-critical V2X traffic and an eMBB agent managing high-throughput services) compete for a constrained pool of shared resources, specifically a total RAN throughput of $R_{total} = 550$~Mbps and an edge CPU capacity of $C_{total} = 100\%$.

Leveraging their respective NDTs as validation sandboxes, the agents execute Monte Carlo simulations ($N=100$) for granular risk assessment grounded in the real-time telemetry provided by DeMAS. To guarantee extreme reliability, the URLLC agent employs a Risk-Aware paradigm targeting a CVaR $CVaR_{99.999}$, whereas the eMBB agent relies on a Risk-Neglect paradigm focusing on average-case performance. 
The decentralised A2A alternating offer protocol is capped at a maximum of 5 rounds. 
The negotiation proposals are continuously evaluated against an internal scoring function that balances service-level compliance with a linear infrastructure energy model factoring in both static overhead and dynamic radio-compute consumption resource factors.

\begin{figure}[t]
	\centering
	\includegraphics[width=1.0\linewidth]{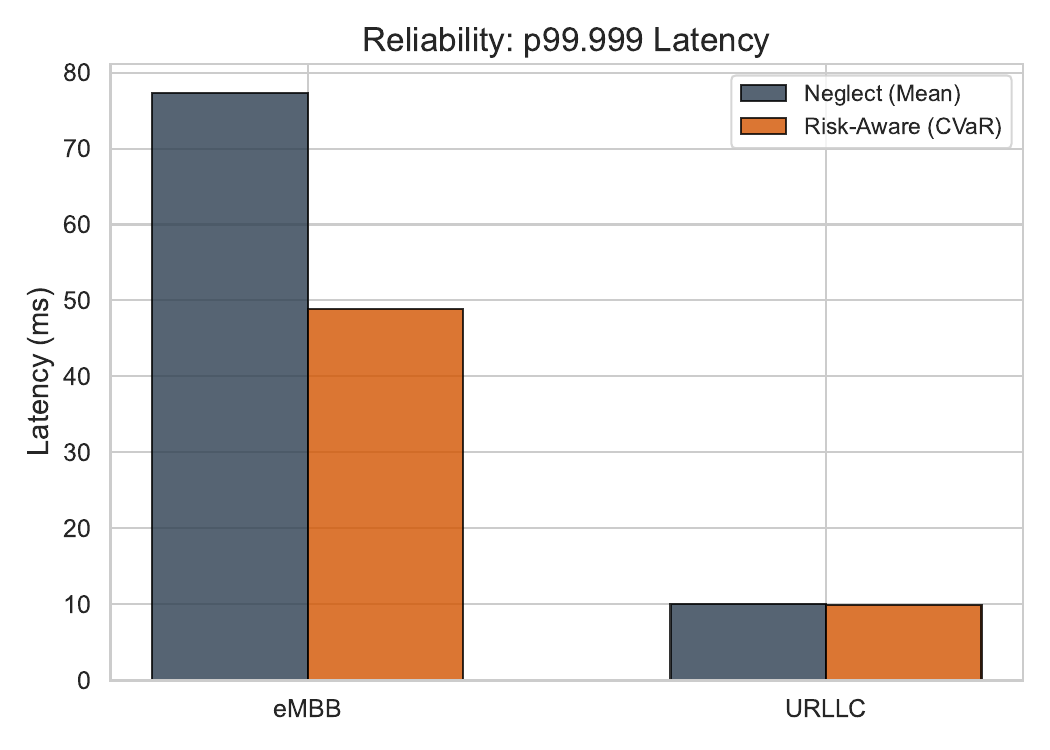}
	\caption{Latency tail distribution}
	\label{fig:tail_latency}
\end{figure}

\begin{figure}[t]
	\centering
	\includegraphics[width=1.0\linewidth]{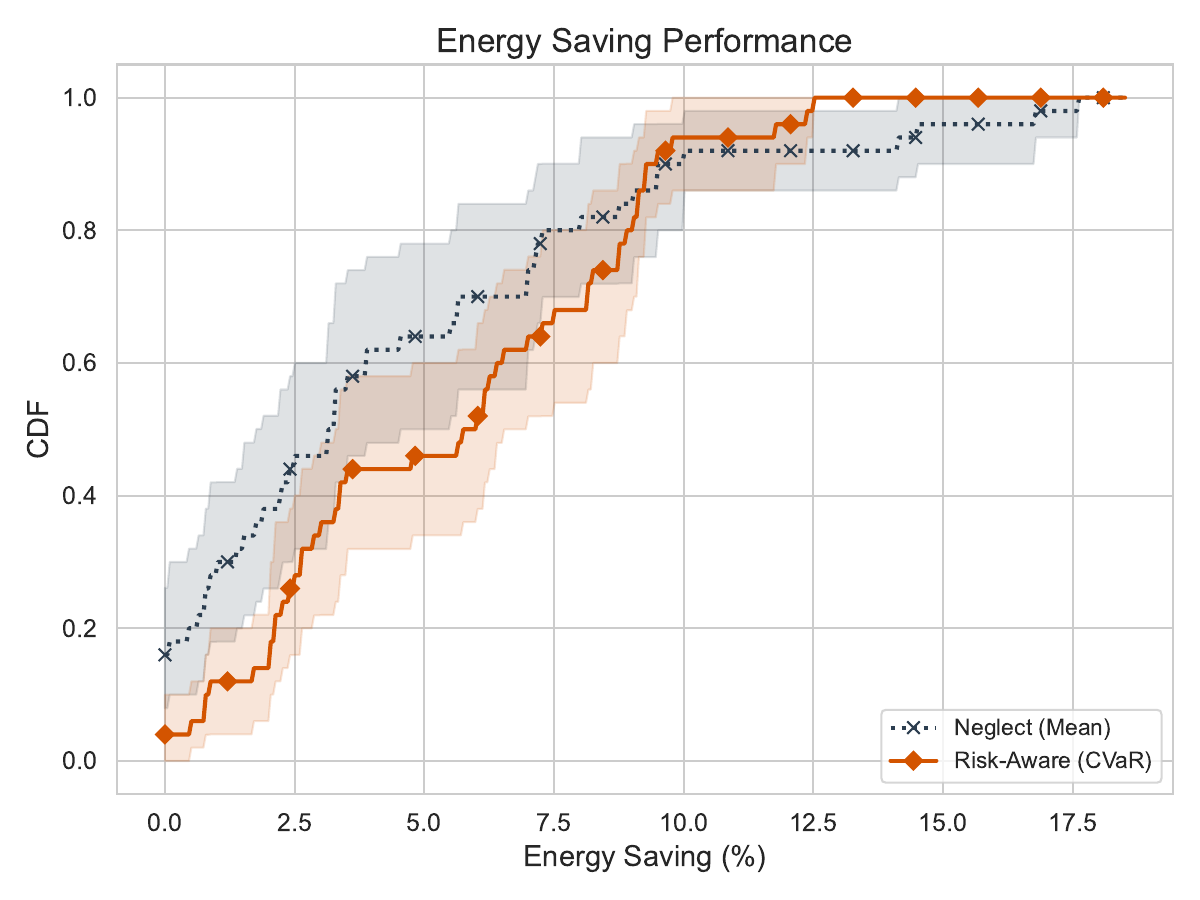}
	\caption{Energy saving distribution for both Mean- and Risk-aware reasoning}
	\label{fig:energy}
\end{figure}


Upon reaching consensus, the API Enforcement Layer executes standard service calls, effectively transforming abstract agreements into physical orchestration by triggering a \texttt{POST /qod/v1/sessions} request to the CAMARA QoD API to enforce radio bandwidth prioritisation, and a \texttt{POST /edge-compute/v1/allocation} call to the MEC Application API for compute resources reservation. The agentic reasoning is powered by the \texttt{gemini-3-flash-preview} model. The environment operates at a temporal granularity of $\tau > 1$~s operating in the non real-time scale which is suitable for network slice management solutions. System constraints define the compute processing rate at 2~Mbps per 1\% CPU utilisation, with specific energy constants set at $C_{bw} = 0.1$~W/Mbps and $C_{cpu} = 0.2$~W/\% to ensure agents prioritise energy efficiency alongside reliability during the final enforcement phase.

\subsection{Performance Analysis}

The framework's efficacy is demonstrated by comparing the Risk-Neglect (Mean) and CVaR paradigms during A2A negotiation. Regarding extreme reliability, the URLLC agent successfully bounds the 99.999th percentile ($p99.999$) latency at the required 10~ms mark across both approaches. Notably, adopting Risk-Aware reasoning yields substantial collateral benefits for the coexisting eMBB service, as shown in Fig.~\ref{fig:tail_latency}, dramatically reducing its extreme tail latency from approximately 77~ms down to 49~ms.

The impact of agentic reasoning on environmental sustainability is captured in Fig.~\ref{fig:energy}, where the Cumulative Distribution Function (CDF) of energy savings reveals that the Risk-Aware approach simultaneously enhances overall system efficiency. Rather than penalising power consumption to guarantee strict reliability limits, the CVaR-targeted paradigm shifts the CDF rightward, improving median (50th percentile) energy savings from roughly 3.5\% under the Neglect baseline to over 5.5\%. These results validate that the framework's internal scoring function effectively harmonises stringent  
SLA enforcement with broader network energy optimisation.


\section{Open research challenges}
\label{sec:next}

The evolution towards agent-driven extensions of the OP for 6G networks not only opens up promising opportunities for fully autonomous operation, but also exposes a set of fundamental technical and operational challenges as detailed below:

\subsubsection{Multi-Agent Decision Stability and Cognitive Biases} As decision-making becomes distributed across interacting agents, LLM-based entities operating over OP interfaces may inherit cognitive biases. During multi-agent coordination, phenomena such as anchoring can constrain the exploration of optimal configurations, while decentralised interactions may trigger premature convergence across federated domains. At scale, these effects heighten the risk of cascading misconfiguration. Addressing these risks requires embedding lightweight mitigation strategies, such as controlled exploration, anchor randomisation, and counterfactual reasoning, directly into OP-enabled workflows. Furthermore, while LLMs provide sophisticated reasoning for complex, multi-variable negotiations, adopting SLMs is critical to balance reasoning depth with the stringent computational and real-time constraints of 6G environments. 

\subsubsection{
Scalable Cross-Domain Observability} Autonomous operation over the OP framework depends on continuous visibility across heterogeneous domains. Although monitoring frameworks such as DeMAS extend the OP with declarative observability, scaling them across federated deployments remains an open challenge. 
Collecting, synchronising, and exposing high-dimensional telemetry across domains may introduce excessive latency and signalling overhead, particularly for time-sensitive closed-loop control. Future work should therefore investigate observability mechanisms that preserve decision-relevant context while limiting cross-domain data exchange through selective sharing, event-driven reporting, or higher-level state abstractions.

\subsubsection{Federated Trust and Operational Privacy} The transition towards federated OP deployments introduces new tensions between collaboration and operational privacy. While OP enables cross-operator service exposure, effective coordination requires partial sharing of network state, which may conflict with competitive constraints, data sovereignty, and internal governance policies. This is particularly relevant in agent-driven OP scenarios, where optimisation decisions depend on cross-domain visibility and multi-agent coordination. Privacy-preserving approaches such as federated learning or parameter-efficient adaptation techniques offer a path forward, enabling coordination without exposing sensitive OP-level data like topology details, traffic patterns, or infrastructure capabilities. 

\subsubsection{Edge-Cloud Native Deployment and State Management} LLM and SLM-based agents introduce non-negligible inference latency, memory pressure and GPU dependencies. Hosting these elements as containerised workloads 
raises new challenges in workload placement, migration and infrastructure scaling across the edge-cloud continuum. Furthermore, the persistent context that agents require for closed-loop reasoning is opposite to the stateless nature of cloud-native deployments, forcing systems to include components such as memory stores and migration-aware servers that preserve continuity across infrastructure restarts, rescheduling, or cross-domain handovers. 

Crucially, the practical deployment of this AI-native paradigm requires reconciling these infrastructure bounds with evolving standardisation and open-source initiatives. Since the proposed architecture builds upon the ETSI SDG OOP, maintaining compatibility with emerging specifications is essential to facilitate real-world adoption across multi-vendor and multi-operator environments. Beyond technical interoperability, widespread acceptance depends on aligning diverse stakeholders with varying operational priorities and business incentives. 
Furthermore, the architecture must remain modular, extensible, and implementation-agnostic to adapt to future 6G requirements and emerging vertical use cases. Achieving such balance between standardisation stability and architectural flexibility is fundamental to long-term sustainability, ecosystem convergence, and industry-wide adoption.

Within this context, the practical realisation of AI-native and agent-driven OP frameworks through platforms such as ETSI SDG OOP introduces an inherent tension between agility and operational control. 
On the one hand, agentic intelligence requires dynamic, extensible OP stacks to integrate evolving AI components, adaptive control loops, and cross-domain runtime negotiations. On the other hand, this flexibility must be reconciled with strict traceability, observability, and conformance to evolving GSMA, CAMARA, and TMF specifications. Specifically, although a vast catalogue of TMF APIs (e.g., for billing and service ordering) is already standardised, these interfaces remain absent from current OP implementations like OOP. Programmatically integrating this business layer directly into the agentic runtime loop is essential for enabling autonomous multi-operator commercial orchestration, yet it introduces strict requirements for runtime governance. These challenges become more pronounced in multi-agent environments, where autonomous entities operate with potentially conflicting policies and must be treated as first-class stakeholders within the system lifecycle. Ensuring that their actions remain auditable, their decision boundaries enforceable, and their behaviour aligned with system-wide constraints introduces non-trivial requirements for runtime governance, policy verification, and accountability. Addressing this tension requires new mechanisms for embedding compliance, monitoring, and control primitives directly into OP implementations.


\section{Conclusion}
\label{sec:conclusion}

This paper presents an architectural evolution of the GSMA OP designed to meet the autonomous and federated demands of 6G while ensuring full alignment with GSMA and CAMARA specifications. By introducing an Agentic-driven Intelligence solution built upon the ETSI SDG OOP reference framework, this work extends the traditional OP into an AI-native orchestration ecosystem that leverages NDTs for grounded predictive reasoning and employs the decentralised A2A negotiation protocol for multi-domain resource coordination. Central to this evolution is the integration of DeMAS, which provides the real-time telemetry and persistent context necessary for autonomous agents to execute closed-loop control across multi-operator domains. 
Our validation scenario confirms that this agentic extension successfully reconciles competing high-priority service requirements via a decentralised multi-agent framework in which specialised URLLC and eMBB entities cooperatively negotiate, guaranteeing strict delay bounds for mission-critical traffic while mitigating tail latency spikes and optimising infrastructure energy efficiency.

Serving as a scalable blueprint for autonomous systems across the federated 6G Cloud–Edge–IoT continuum, realising its full potential requires addressing fundamental research challenges, including key operational factors such as mitigating cognitive biases within distributed agent reasoning and ensuring operational privacy in federated environments. By maintaining alignment with emerging open-source implementations and addressing these open challenges, the proposed architecture paves the way for a truly programmable, intelligent, and collaborative multi-operator ecosystem.

\bibliographystyle{IEEEtran}
\bibliography{references}

\end{document}